\documentclass[12pt]{iopart}
\usepackage{cite}
\usepackage{indentfirst, iopams, placeins, graphicx, caption}
\usepackage[colorlinks=true,allcolors=blue,bookmarks=false,hypertexnames=true]{hyperref}
\begin{document}
\title[Energy transport in $\mathbb{Z}_3$ chiral clock model]{Energy transport in $\mathbb{Z}_3$ chiral clock model}
\author{Naveen Nishad \& G J Sreejith}
\address{Indian Institute of Science Education and Research, Pune 411008 India}
\vspace{10pt}

\begin{abstract}

We characterize the energy transport in a one dimensional $\mathbb{Z}_3$ chiral clock model. The model generalizes the $\mathbb{Z}_2$ symmetric transverse field Ising model (TFIM).
The model is parametrized by a chirality parameter $\theta$, in addition to $f$ and $J$ which are analogous to the transverse field and the nearest neighbour spin coupling in the TFIM. Unlike the well studied TFIM and XYZ models, does not transform to a fermionic system.
We use a matrix product states implementation of the Lindblad master equation to obtain the non-equilibrium steady state (NESS) in systems of sizes up to $48$.
We present the estimated NESS current and its scaling exponent $\gamma$ as a function of $\theta$ at different $f/J$. 
The estimated $\gamma(f/J,\theta)$ point to a ballistic energy transport along a line of integrable points $f=J\cos{3\theta}$ in the parameter space; all other points deviate from ballistic transport. Analysis of finite size effects within the available system sizes suggest a diffusive behavior away from the integrable points.

\end{abstract}

\maketitle
\section{Introduction}
Though energy transport has been studied for a long time, a microscopic description of energy transport in interacting quantum and classical systems is still under development, with many recent insights on connections between chaos and transport aided by the improved simulation methods.
In classical systems, chaos is neither necessary nor a sufficient condition\cite{Dhar} for diffusive transport. Fermi-Pasta-Ulam problem has a positive Lyapounov exponent, but does not exhibit diffusive heat conduction in any parameter regime.

An extensive amount of work on high temperature transport focusing on spin-half models in one dimensional (1D) quantum systems \cite{Saito_2003,Mej_a_Monasterio_2005,Heidrich-Meisner2007,Saito2,Saito1996ThermalCI,PhysRevB.55.11029,PhysRevB.98.180201,PhysRevLett.117.040601,Sun_2010,PhysRevB.102.184304} have shown that breaking integrability generally leads to diffusive  energy transport. It has been analytically argued that integrability in clean systems typically leads to ballistic energy transport \cite{PhysRevLett.74.972,PhysRevB.55.11029}. Interestingly, the relation does not extend to other conserved currents \cite{Prosen_2009,PhysRevLett.122.127202,PhysRevLett.106.220601, Ljubotina2017,PhysRevB.98.235128}. 
The XXZ chain in its zero-magnetization sector shows ballistic energy transport in all phases but spin transport is ballistic in the easy plane phase, diffusive in the easy axis phase, and super-diffusive at the isotropic point.\cite{PhysRevLett.106.220601,PhysRevB.98.180201}. Both spin and energy transport are found to be ballistic in other magnetisation sectors.\cite{PhysRevB.55.11029,PhysRevE.91.042129,PhysRevB.99.094435} 
On the other hand the same model, with a local longitudinal field, is non-integrable but shows ballistic spin transport\cite{PhysRevB.98.235128}. Disorder further enriches transport physics in such systems \cite{PhysRevB.99.094435}.

In this work, we step away from the well-studied spin-1/2 model and explore a model with a three dimensional local Hilbert space, namely the $\mathbb{Z}_3$ symmetric chiral clock chain\cite{PhysRevB.24.398,PhysRevB.24.5180,HOWES1983169} which generalizes of the $\mathbb{Z}_2$ symmetric TFIM\cite{PFEUTY197079}.
The latter which is mappable to free fermions is integrable and exhibits ballistic energy transport \cite{PhysRevB.83.214416}. 
The $\mathbb{Z}_3$ clock model Hamiltonian is integrable in a fine tuned set of parameters but not in general. While the model shares several features with the TFIM, it is not mappable to a free fermionic Hamiltonian. We aim to address the question of how energy transport is affected by the model parameters, in particular how integrability affects transport in this model. Transport through the chain is simulated using the Lindblad master equation (LME) approach implemented using matrix product state (MPS) techniques \cite{Prosen_2009,Mendoza_Arenas_2013,PhysRevLett.106.220601,PhysRevB.102.184304}.

Our paper is structured as follows.
In Sec. \ref{model}, we describe the chiral clock model and present the details of the Lindblad dissipators. 
We then describe the details for the MPS implementation of the LME in Sec. \ref{Numerical Implementation}. 
We find that under a change of basis, the LME and transport properties in one part of the parameter space can be related to that in another part, reducing the parameter space to be studied. This is described in Sec. \ref{Symmteries}. 
Results for the simulations are presented in the Sec. \ref{Ferromagnetic} and conclude with Sec. \ref{Conclusion}.

\section{Model}\label{model}

The $\mathbb{Z}_{3}$ chiral clock model for a chain of $N$ spins in 1D, is described by the Hamiltonian\cite{PhysRevB.24.398,PhysRevB.24.5180,HOWES1983169,Fendley2012}
\begin{equation}
H(\theta,\phi)=-Je^{\iota\theta}\sum_{i=1}^{N-1}\sigma_i\sigma_{i+1}^\dagger-fe^{\iota\phi}\sum_{i=1}^N\tau_i+\rm{H.c.}
\label{eq:bareChiralHamiltonian}
\end{equation}
Each spin has a three dimensional Hilbert space, and the local operators $\sigma$ and $\tau$ have the following matrix representation
\begin{equation}
	\sigma=\left(\begin{array}{ccc}
	1 & 0 & 0\\
	0 & \omega & 0\\
	0 & 0 & \bar{\omega}
	\end{array}\right)\;\;\;\tau=\left(\begin{array}{ccc}
	0 & 1 & 0\\
	0 & 0 & 1\\
	1 & 0 & 0
	\end{array}\right)
\end{equation}
where $\omega=\exp(2\pi\iota/3)$. We will represent the single site eigenstates of the $\sigma$ operator as $\left|1\right \rangle$, $\left|\omega\right \rangle$ and $\left|\bar{\omega}\right \rangle$.
Operators $\sigma$ and $\tau$ satisfy the algebra  $\sigma_{i}^{3}=\tau_{i}^{3}=1$, $\sigma_{i}\tau_{i}=\bar{\omega}\tau_{i}\sigma_{i}$, and $\sigma_{i}\tau_{j}=\tau_{j}\sigma_{i}$ for $i\neq j$. This algebra is a $\mathbb{Z}_3$ analog of the algebra of Pauli matrices $\sigma_z$ and $\sigma_x$. Interplay between $f,\;\theta$ and $\phi$ results in a rich ground state phase diagram\cite{PhysRevB.92.035154,PhysRevA.98.023614,PhysRevB.98.205118} hosting trivial, topological and incommensurate phases.

The model has a global $\mathbb{Z}_3$ parity symmetry associated with the operator $\mathcal{P}=\Pi_i\tau_i$. Apart from the global parity symmetry, the model can have other symmetries\cite{Mong_2014} namely time reversal $\mathcal{T}$, charge conjugation $\mathcal{C}$, and spatial inversion $\mathcal{S}$ depending on the values of parameters $\theta$ and $\phi$. 
Under these symmetry transformations, $\sigma$ and $\tau$ operators transform as $\mathcal{T}^\dagger \sigma\mathcal{T}=\sigma^\dagger$, $\mathcal{T}^\dagger \tau\mathcal{T}=\tau$, $\mathcal{C}^\dagger \sigma\mathcal{C}=\sigma^\dagger$, and $\mathcal{C}^\dagger \tau\mathcal{C}=\tau^{\dagger}$. Charge conjugation swaps the states $|\omega\rangle$ and $|\bar{\omega}\rangle$. Spatial inversion changes site index $i\to N-i+1$. All three symmetries are present at $\theta=\phi=0$ while the model has only spatial inversion symmetry when $\theta= 0$ and $\phi\neq0$. None of the three symmetries are present when both $\theta$ and $\phi$ are non-zero. 
In this work we will focus on the models with $\phi=0$ for simplicity. For $\phi=0$ and $\theta\neq 0$, the individual symmetries $\mathcal{C}$ and $\mathcal{S}$ are broken but their products are preserved.

At $f=0$, all the eigenstates of Hamiltonian can be chosen to be direct products of eigenstates of $\sigma_i$. Energy of each eigenstate is $-2J\sum_i\cos(\theta+\alpha_i)$, where $\alpha_i={\arg}(\langle\sigma_i\rangle/\langle\sigma_{i+1}\rangle)$ which take values from $\{0,\pm 2\pi/3\}$.
When $\theta\in(-\pi/3,\pi/3)$, all the spins in ground state are aligned in the same direction, either in $1,\;\omega$ or $\bar{\omega}$. Ground state for $\theta\in (\pm\pi/3,\pm\pi)$ has consecutive spins oriented at relative angle of $\pm2\pi/3$. Parameter $f$ tunes quantum fluctuation in the model. 
At large $f$, the ordered phase is destroyed forming a paramagnetic phase. 
A second order phase transition separates the $\mathbb{Z}_3$ symmetry broken phase (small $f$) and $\mathbb{Z}_3$ symmetric phase(large $f$). The model was shown to be integrable along the line $f=J\cos{3\theta}$ inside the ordered phase\cite{AUYANG1987219}. 

There has been limited studies of transport properties in the model. Non-equilibrium current in $\mathbb{Z}_3$ chiral clock chain with alternating sites are different temperatures have been studied in Ref \cite{puebla2021open}. At the critical integrable point described by $f=J$ and $\theta=0$, energy transport between a ground state and high energy state was studied in a generalized hydrodynamics framework in Ref. \cite{CCMGHD}. We will study the energy transport in the ferromagnetic ($f<J$) regime and at varying values of $\theta$. 

A natural framework for investigation of transport properties is to attach baths with different characteristic temperatures at the opposite ends of the chain.
This temperature difference creates an energy gradient and energy flow from high to low temperature end. 
In Ref.\cite{Prosen_2009}, Prosen $et\;al.$ introduced the idea of using few-site jump operators to study transport properties under the dissipative dynamics of LME. This strategy provides computational simplicity and speedup leading to its extensive use for studying spin and fermionic chains\cite{PhysRevB.98.235128,PhysRevLett.106.220601,PhysRevB.99.094435,PhysRevB.102.195142,Mendoza_Arenas_2013,PhysRevB.102.184304,PhysRevLett.117.040601,PhysRevB.86.125111}.
It has been argued that the local Lindblad approximations cannot faithfully  reproduce the coherences produced by coupling to an actual quantum environment \cite{ManasLME}. The local Lindblad operators we use are intended to maintain local energy densities at the ends of the chain rather than to mimic a realistic quantum bath. We assume that the transport properties are independent of the manner in which the local energy density is realized.

\begin{figure}
\centering
\includegraphics[width=0.9\columnwidth]{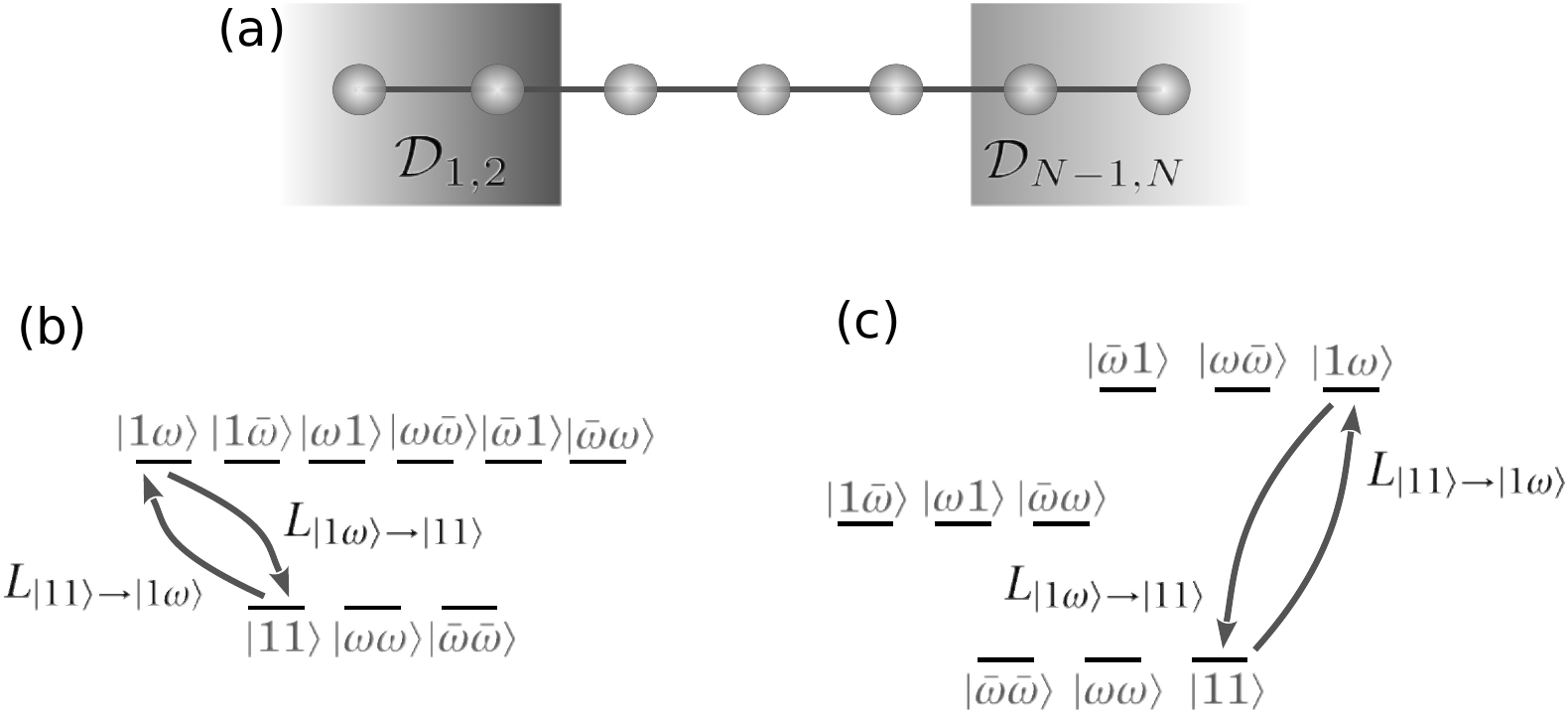}
\caption{Pictorial representation of the bath system setup is shown in panel (a). Action of the jump operators $L_{|11\rangle\to|1\omega\rangle}$ and $L_{|1\omega\rangle\to|11\rangle}$ are shown for the ferromagnetic regime for dissipators $\mathcal{D}^0$ and $\mathcal{D}^\theta$ is shown in (b) and (c).}  \label{fig:Boundary_dissipators}
\end{figure}

Dissipative dynamics of the system with bath attached at both ends is given by the LME\cite{Lindblad1976}
\begin{equation}
	\partial_{t}\rho(t)=\iota[\rho(t),H]+\mathcal{D}[\rho (t)]
	\label{eq:LME}
\end{equation}
where $\rho$ is the density matrix of the system and $\mathcal{D}[\rho]$ is the Lindblad dissipator. The dissipator acts on the two sites at each end of the chain
\begin{equation}
\mathcal{D}[\rho]=\mathcal{D}_{1,2}(\beta_L)[\rho]+\mathcal{D}_{N-1,N}(\beta_R)[\rho]
\end{equation}
where $\beta_L$ and $\beta_R$ parametrize the inverse temperature for left and right end of the chain respectively. 
We define two site boundary dissipative term $\mathcal{D}_{i,j}(\beta)[\rho]$ acting on the spin at site $i$ and $j$ using jump operators $L_{a\to b}=|b\rangle\langle a|$ as
\begin{equation}
    \mathcal{D}_{i,j}(\beta)[\rho]=\lambda\sum_{ab}\Gamma_{+}(\beta)[L_{a\to b},\rho L_{b\to a}]+\Gamma_{-}(\beta)[L_{a\to b},\rho L_{b\to a}]^{\dagger}
    \label{Dissipatordetails}
\end{equation}
Here $\lambda$ quantifies the coupling strength between the system and the bath. The two-site states $|a\rangle$ and $|b\rangle$ are the eigenstates, with energy eigenvalues $E_a$ and $E_b$, of a two-site Hamiltonian $h_{i,j}$ acting on sites $i$ and $j$.
The transition rates are given by $\Gamma_{\pm}=e^{\mp\beta (E_b-E_a)/2}$ as shown in Fig. \ref{fig:Boundary_dissipators}(b). $h_{i,j}$ contains the dominant terms of the full Hamiltonian restricted to the ends of the chain. 

In this manuscript, we have used two types of local boundary dissipators denoted by $\mathcal{D}_{i,j}^0$ and $\mathcal{D}_{i,j}^\theta$, constructed using two different choice of the two-site Hamiltonians $h_{i,j}^0$ and $h_{i,j}^\theta$. 
\begin{enumerate}
\item 
$\mathcal{D}_{i,j}^0$ is defined using the two-site Hamiltonian $h_{i,j}^0=-J\sigma_i\sigma_j^{\dagger}+\rm{H.c.}$ 
The ground state of $h^0$ is three fold degenerate ($\left|11\right \rangle$, $\left|\omega\omega\right\rangle$, and $\left|\bar{\omega}\bar{\omega}\right \rangle$) and its excited state is six-fold degenerate with an energy gap of $3J$ between them. We have included Lindblad jump operators only between the non-degenerate eigenstates of $h^0$.
We note that due to the ferromagnetic nature of $h^0$, use of $\mathcal{D}^0_{i,j}$ makes sense only when $\theta\in(-\pi/3,\pi/3)$ where the ferromagnetic states have a lower energy.
\item
The dissipator $\mathcal{D}_{i,j}^\theta$ is constructed using $h_{i,j}^\theta=-Je^{\iota\theta}\sigma_i\sigma_j^\dagger+\rm{H.c}$.
The ground state of $h^\theta$ is still three fold degenerate. 
These are the ferromagnetically aligned states when $\theta\in(-\pi/3,\pi/3)$ and 
When $\theta\in(\pi/3,\pi)$ the ground states are $\left|1\omega\right\rangle$, $\left|\omega\bar{\omega}\right\rangle$, and $\left|\bar{\omega}1\right\rangle$.
Similarly, when $\theta\in(-\pi,-\pi/3)$, $\left|1\bar{\omega}\right\rangle$, $\left|\bar{\omega}\omega\right\rangle$, and $\left|\omega 1\right\rangle$ are the ground states.
Introduction of the $e^{\iota\theta}$ prefactor in the local Hamiltonian breaks the six fold degeneracy of the excited states (except when $\theta$ is a multiple of $2\pi/3$). In defining the dissipator, we have included transitions between degenerate states of $h^\theta$.
\end{enumerate}
A schematic representation of the jump operators in $\mathcal{D}_{i,j}^0$ and $\mathcal{D}_{i,j}^\theta$ are shown in Fig.  \ref{fig:Boundary_dissipators}(b) and Fig. \ref{fig:Boundary_dissipators}(c) respectively. 
The effective local temperatures generated by the two different dissipators as well as the length scales for thermalization near the boundary will be different for the two choice of dissipators. However we expect that qualitative features of transport will be similar in the two cases if the results are independent of the precise form of the bath. We indeed find this to be the case.

 
For finite dimensional systems, the LME has at least one fixed point (See Sec 4.2.2 of Ref. \cite{watrous_2018}). In small  systems of upto 5 sites, we diagonalized the Liouvillian and found that it has a unique $0$-eigenvalue state. Assuming the uniqueness to be true in larger systems, the time evolution under the above LME should approach a unique non-equilibrium steady state (NESS) defined as
\begin{equation}
    \rho_{\rm{NESS}}^\theta=\lim_{t\to\infty}\rho (t)
\end{equation}
To obtain the NESS, we integrated the LME till large $t$ and used saturation of local observables - energy current, energy density and magnetization on each site to check approach to steady state.

The local energy density $E_i^\theta$ at site $i$ is chosen to be the three site operator
\begin{equation}
	E_i^\theta=-\frac{J}{2} e^{\iota\theta}(\sigma_{i-1}\sigma_i^\dagger+\sigma_i\sigma_{i+1}^\dagger)-f\tau_i+\rm{H.c}
\end{equation} 
The current operator on the bond between sites $i$ and $i+1$ can be written as $I_i^\theta=\iota[E_{i+1}^\theta,E_{i}^\theta]$. We evaluate this to be 
 \begin{equation}
     I_i^\theta=\iota\frac{fJe^{\iota\theta}}{2}(I_i^{(1)}+I^{(2)}_i)+\rm{H.c}
 \end{equation}
where
 \begin{eqnarray}
     I_i^{(1)} &= (\omega-1)\sigma_i(\tau_i+\tau_{i+1})\sigma_{i+1}^\dagger\nonumber\\
     I_i^{(2)} &= (\bar{\omega}-1)\sigma_i(\tau_i^\dagger+\tau_{i+1}^\dagger)\sigma_{i+1}^\dagger\nonumber
\end{eqnarray}
The energy and current operators satisfy the discrete continuity equation 
\begin{equation}
\frac{d E_i^\theta}{dt}=\iota[H(\theta,0),E_i^\theta]=I_i^\theta-I_{i-1}^\theta
\end{equation}

The expectation value $\langle I_i\rangle_{\rm{thermal}}={\rm Tr}(e^{-\beta H}I_i)/\mathcal{Z}=0$ of the chosen form of the current operator is zero in the thermal state.
This can be seen as follows.
It can be checked that the unitary symmetry transformation operator $\mathcal{C}\mathcal{S}$ introduced in Sec. \ref{model} commutes with the Hamiltonian $H(\theta,0)$ and anticommutes with $I^\theta$.
Now consider the expectation value of the symmetry transformed current:
\begin{eqnarray}
{\rm Tr} [e^{-\beta H}(\mathcal{CS})^{\dagger} I^\theta (\mathcal{CS})]=\langle I^\theta\rangle_{\rm thermal} \nonumber\\ {\rm Tr} [e^{-\beta H}(\mathcal{CS})^{\dagger} I^\theta (\mathcal{CS})]= -\langle I^\theta\rangle_{\rm thermal}
\end{eqnarray}
suggesting that the current as defined is zero in the the thermal state.
In the first equality we have used the cyclic property of the trace and the commutation of $\mathcal{CS}$ with $H$. In the second equality, we have used the anticommutation property with $I$.

Fick's law can be generalized to all transport regimes using an empirical exponent $\gamma$ as
\begin{equation}
\langle I\rangle=\kappa\times(\langle E_N\rangle - \langle E_1 \rangle)
\end{equation}
where $\kappa$ is steady state energy conductance which scales as $1/N^\gamma$ with system size $N$. Ballistic and diffusive transport are characterized by $\gamma=0$ and $1$ respectively. Systems exhibiting a conduction with $0<\gamma <1$ and $\gamma >1$ are said to have super-diffusive and sub-diffusive transport. We characterize the transport in the clock model from the scaling of $\langle I \rangle$ with $N$ allowing us to estimate the exponent $\gamma$.

\section{NESS currents at $\theta$, $\theta+2\pi/3$ and $-\theta$}\label{Symmteries}
In this section, we show that, under the time evolution (Eq. \ref{eq:LME}) with the dissipator $\mathcal{D}^\theta$, the NESS current at $\theta$ is same as that at $-\theta$ and $\theta+2\pi/3$. Using this equivalence of transport behavior at different $\theta$, we can reduce the parameter region to be studied from $\theta\in[0,2\pi)$ to $[0,\pi/3]$.
To see the equivalence, we consider the unitary operators $\mathcal{U}_1=\Pi_i\tau_i^i$ and $\mathcal{U}_2=\Pi_i\mathcal{C}_i$. These transform the Hamiltonian as follows
\begin{eqnarray}
    \mathcal{U}_1^\dagger H(\theta,0)\mathcal{U}_1=H(\theta+2\pi/3,0)\label{eq:2piby3symmetry}\nonumber\\
    \mathcal{U}_2^\dagger H(\theta,0)\mathcal{U}_2=H(-\theta,0)\label{eq:minusthetasymmetry}
\end{eqnarray} 
Transformation of the dissipator $\mathcal{D}^\theta[\rho]$ under the unitaries $\mathcal{U}_1$ and $\mathcal{U}_2$ is given by 
\begin{eqnarray}
    \mathcal{U}_1^\dagger\mathcal{D}^\theta[\rho]\mathcal{U}_1=\mathcal{D}^{\theta +2\pi/3}[\mathcal{U}_1^\dagger\rho\mathcal{U}_1]\label{eq:D2piby3symmetry}\nonumber\\
    \mathcal{U}_2^\dagger\mathcal{D}^\theta[\rho]\mathcal{U}_2=\mathcal{D}^{-\theta}[\mathcal{U}_2^\dagger\rho\mathcal{U}_2]\label{eq:Dminusthetasymmetry}
\end{eqnarray}
With these, it can be checked that the stationary solution $\rho_{\rm NESS}$ to the LME (Eq. \ref{eq:LME}) at $\theta+2\pi/3$ and at $-\theta$ are related to the solution at $\theta$ by
\begin{eqnarray}
    \mathcal{U}_1^\dagger \rho_{\rm{NESS}}^\theta\mathcal{U}_1=\rho_{\rm{NESS}}^{\theta+2\pi/3}\nonumber\\
    \mathcal{U}_2^\dagger \rho_{\rm{NESS}}^\theta\mathcal{U}_2=\rho_{\rm{NESS}}^{-\theta}
\end{eqnarray}
Note that we have implicitly assumed that there is only one NESS at each $\theta$.
The energy density $E_i^\theta$ and current $I_i^\theta$ transform similarly to $H(\theta,0)$ under $\mathcal{U}_1$ and $\mathcal{U}_2$.

\begin{figure}
    \centering
 	\includegraphics[width=0.55\columnwidth]{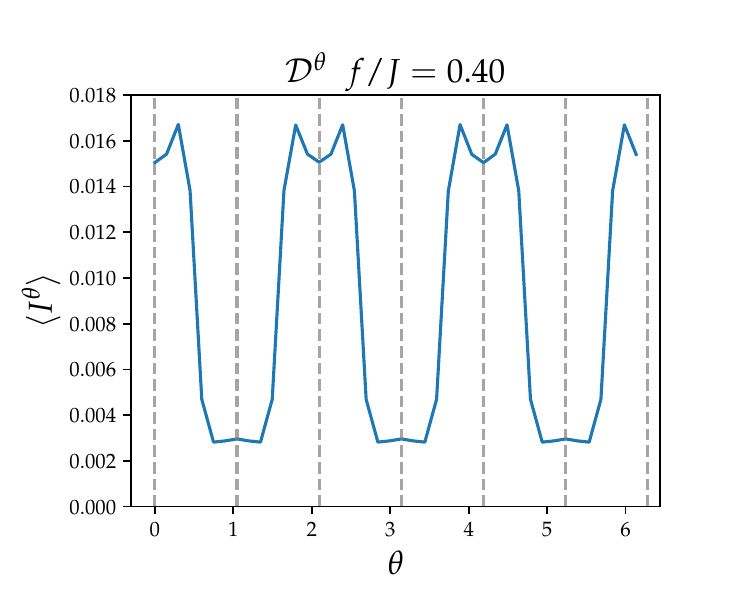}
 	\caption{Plot of the NESS current $I^\theta$ as a function of $\theta$ for its full range of values from $0$ to $2\pi$ showing the equivalence of transport properties at $\theta$, $-\theta$ and $\theta+2\pi/3$. Data is shown for the system size $N=14$, with the dissipator $\mathcal{D}^\theta$ at model parameters $f/J=0.4$. Vertical lines show multiples of $\pi/3$.
 	\label{fig:scan_theta}}
 \end{figure}

The thermal expectation value of the current at $-\theta$ is given by
\begin{equation}
\langle I^{-\theta} \rangle ={\rm Tr} [\rho^{-\theta} I^{-\theta} ]={\rm Tr} [\mathcal{U}_2^\dagger\rho^{\theta} \mathcal{U}_2  \mathcal{U}_2^\dagger I^{\theta} \mathcal{U}_2 ] = {\rm Tr} [\rho^{\theta} I^{\theta} ] = \langle I^{\theta} \rangle
\end{equation}
Similarly, we find that $\langle E_i^\theta\rangle=\langle E_i^{-\theta}\rangle=\langle E_i^{\theta+2\pi/3}\rangle$ and $\langle I_i^\theta\rangle=\langle I_i^{-\theta}\rangle=\langle I_i^{\theta+2\pi/3}\rangle$.
These symmetries in the current and energy as a function of $\theta$ were verified in our numerical implementation of the LME. In Fig. \ref{fig:scan_theta} symmetry in NESS current is shown for system size $N=14$ and $f/J=0.4$ using the dissipator $\mathcal{D}^\theta$. These results allow us to use the transport properties evaluated in $\theta\in[0,\pi/3]$ to infer the transport properties in the whole range $[0,2\pi]$.

Similar arguments for the case of the dissipator $\mathcal{D}^0$ shows that the energy and currents at $\theta$ and $-\theta$ are equal to each other. 

\section{Numerical Implementation}\label{Numerical Implementation}
Evolution under the LME (Eq. \ref{eq:LME}) was implemented using the Matrix Product State (MPS) formalism where we represent $\rho$ as an MPS of the form 
\begin{equation}
|\rho\rangle=\sum_{\sigma,\sigma '}A^{\sigma_{1}\sigma_{1}^{'}}A^{\sigma_{2}\sigma_{2}^{'}}\ldots A^{\sigma_{N}\sigma_{N}^{'}}|\sigma_{1}\sigma_{2}\ldots\sigma_{N}\rangle|\sigma_{1}^{'}\sigma_{2}^{'}\ldots\sigma_{N}^{'}\rangle
\end{equation}
Each tensor $A$ has physical indices of dimension 9. 
The MPS is normalized such that the density matrix satisfies the trace preserving condition:
\begin{equation}
    \sum_{\sigma\sigma '}A^{\sigma_1\sigma_1^{'}}A^{\sigma_2\sigma_2^{'}}\ldots A^{\sigma_N\sigma_N^{'}}\delta_{\sigma_1\sigma_1^{'}}\delta_{\sigma_2\sigma_2^{'}}\ldots\delta_{\sigma_N\sigma_N^{'}}=1
\end{equation}
In the LME (Eq. \ref{eq:LME}) operators can act on the density matrix $\rho$ either from the left or right. Equivalent matrix product operator for the right and left action of operators on $\rho$ contracts with non primed and primed indices respectively. We can write Eq. \ref{eq:LME} in the super-operator form 
\begin{equation}
	\partial_{t}|\rho\rangle=\hat{\mathcal{L}}|\rho\rangle
	\label{Liouvillian}
\end{equation}
where $\hat{\mathcal{L}}$ is a time independent super-operator called Liouvillian operator. The solution to Eq. \ref{Liouvillian} which can be formally written as $|\rho(t)\rangle=e^{{\hat{\mathcal{L}}}t}|\rho(0)\rangle$ can be evaluated using a fourth-order approximant to MPO similar to those used in Refs. \cite{PhysRevB.96.195117,PhysRevB.91.165112}. Matrix exponential approximant of any order can be expressed as product of several first order approximants $W^{{\rm II}}(\tau)=\mathbb{I}+\tau\mathcal{L}$ as
\begin{equation}
    W^{{\rm II}}(\tau_{1})W^{{\rm II}}(\tau_{2})...W^{{\rm II}}(\tau_{n})={\rm exp}(\mathcal{L}t)+\mathcal{O}(t^{p+1})
    \label{Fourthorder}
\end{equation}
where $\tau$'s are complex numbers proportional to $t$. To obtain an approximant correct till order $p$, we match coefficients of $t$ of each order up to $p$ on both sides of Eq. \ref{Fourthorder}. The $\tau_i$ are solutions of these $p$ simultaneous nonlinear equations.

Assuming that the fixed point is unique, the choice of initial state should not affect the NESS. For completeness we describe the initial state preparation. We started with an infinite temperature state and time evolved it under the following Liouvillian $\mathcal{L}'[\rho]$
\begin{equation}
    \mathcal{L}'[\rho]\equiv \iota[H,\rho]+\sum_{i=1}^{N-1}\mathcal{D}_{i,i+1}(\beta_{i})[\rho]
\end{equation}
with Lindblad dissipators $\mathcal{D}_{i,i+1}$ (Eq. \ref{Dissipatordetails}) connected to all sites with inverse temperature at each site linearly varying with site number between $\beta_L$ and $\beta_R$. The steady state of the time evolution under $\mathcal{L}'(t)$ is later used as an initial state for the actual time evolution. 
The initial state as well as the time evolved states are in equal mixtures of the three $\mathbb{Z}_3$ parity quantum numbers.

The inverse temperatures at the left and right ends are $\beta_L=0.133$ and $\beta_R=0.266$ respectively. The spin coupling is set to be $J=1$ and the coupling to the Lindblad dissipators is set to $\lambda=0.05$. Simulations were performed for systems with $f=0.4$ and for a set of $\theta$ in the range $[0,\pi/3]$. Calculations were separately performed using the two different Lindblad dissipators $\mathcal{D}^0$ and $\mathcal{D}^\theta$. In all of our calculations, bond dimension $\chi$ being used is 200. For a select set of parameters we increased the bond dimension to 800, and no significant change was observed beyond 200 in the local observables. 
\section{Results}\label{Ferromagnetic}
In this section we report the main results of the numerical simulations. The estimated current and energy density in the NESS, and the scaling exponent $\gamma$ of the current as a function of system size are presented. In addition, we also present the level spacing statistics and the operator space entanglement entropy in the NESS.
\subsection{NESS Current and Conductance}
\begin{figure}
    \centering
    \includegraphics[width=\columnwidth]{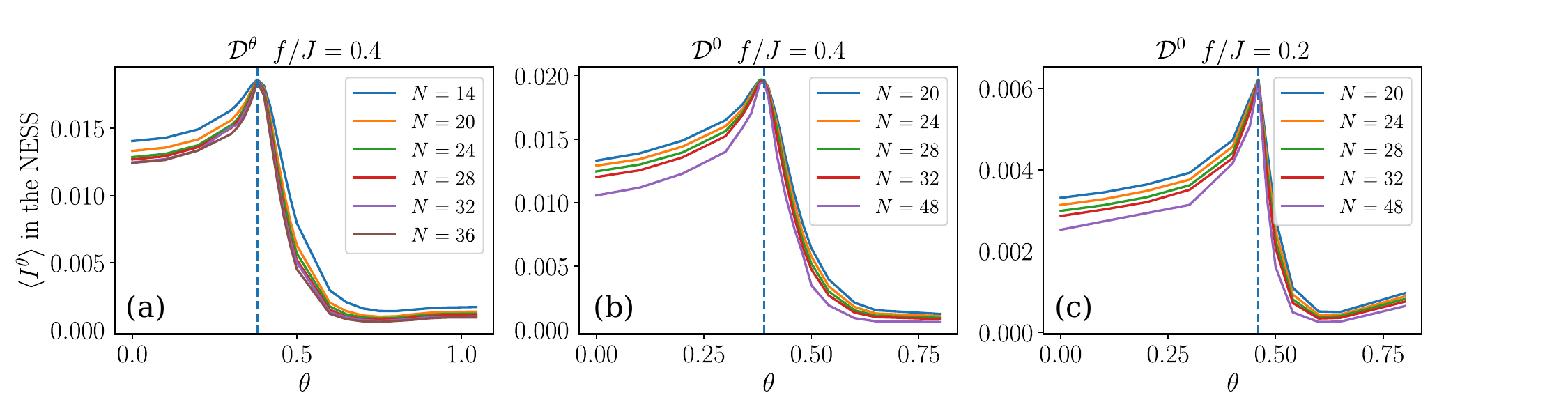}
    \caption{NESS current $\langle I^\theta\rangle$ as a function of $\theta$ in the ferromagnetic regime. 
    Panel (a) shows the current when the Liouvillian is defined using the dissipator $\mathcal{D}^\theta$ and $f/J=0.4$. 
    Panel (b) and (c) show the current for the case of the dissipator $\mathcal{D}^0$ with $f/J=0.4$ and $f/J=0.2$ respectively.
 Different lines indicate different system sizes. The peak current appears at the integrable point $\theta=\cos^{-1}(f/J)/3$ shown by vertical dashed lines in all cases.}\label{fig:Current_vs_theta}
\end{figure}

\begin{figure}
    \centering
    \includegraphics[width=\columnwidth]{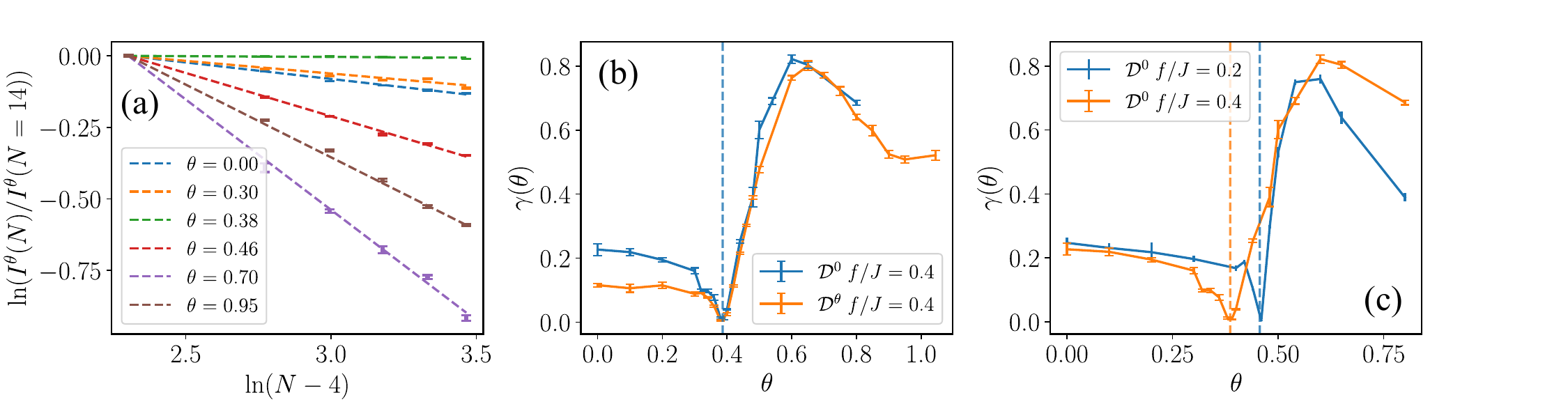}
    \caption{Panel (a) presents $I^\theta (N)$ (rescaled by the current in the smallest system size) vs  $N-4$ in log-scale for fixed values of $\theta$ for $f/J=0.4$ and using the dissipator $\mathcal{D}^\theta$. Panels (b) and (c) show the scaling exponent $\gamma$ estimated from the system size dependence of the NESS current $I^\theta(N)$. Estimated $\gamma$ is plotted as function of $\theta$ in panels (b) and (c). In panel (b), we compare the exponent $\gamma$ obtained from the two different choice of dissipators $\mathcal{D}^0$ and $\mathcal{D}^\theta$. In panel (c) we compare the estimated $\gamma$  obtained using the same dissipator $\mathcal{D}^0$ but for  $f/J=0.2$ and $f/J=0.4$.}\label{fig:Gamma_ferro}
\end{figure}
The mean NESS energy current $\langle I^\theta\rangle=\sum_i \langle I^\theta_i\rangle/N$ as a function of the chiral parameter $\theta$ is shown in Fig $\ref{fig:Current_vs_theta}$ (results do not change if the current at the center of the chain is used instead)
Panel (a) shows the NESS current obtained using the dissipator $\mathcal{D}^\theta$ for parameter $f/J=0.4$. 
Panels (b) and (c) show the same for the dissipator $\mathcal{D}^0$ for model parameter $f/J=0.4$ and $0.2$ respectively. 
In all cases we find a peak current at the $\theta$ where we expect the system to be integrable. 
When the model parameters are changed from $f/J=0.4$ to $f/J=0.2$, the $\theta$ at which the model is integrable changes. Accordingly the location of the peak current also changes.
The NESS current is independent of the system size at the integrable point, consistent with it exhibiting a ballistic transport.
The current decreases with the system size at other $\theta$. These qualitative features are the same for both choice of dissipators.

At each value of $\theta$, the system size dependence of the NESS current can be parametrized using $\gamma$ obtained by fitting the NESS current measured in different system of sizes from $N=14$ to $N=48$ to the form $I^\theta(N_{\rm{eff}})=AN_{\rm{eff}}^{-\gamma(\theta)}$. $N_{\rm{eff}}$ is the effective length of the chain which is $N-4$ as two spins from each end is associated with the Lindblad dissipators. Figure \ref{fig:Gamma_ferro}(a) shows the current as a function of system size for a representative set of values of $\theta$. Within the range of system sizes accessible, we are able to fit the data to the power law form.

\begin{figure}
    \centering
 	\includegraphics[width=\columnwidth]{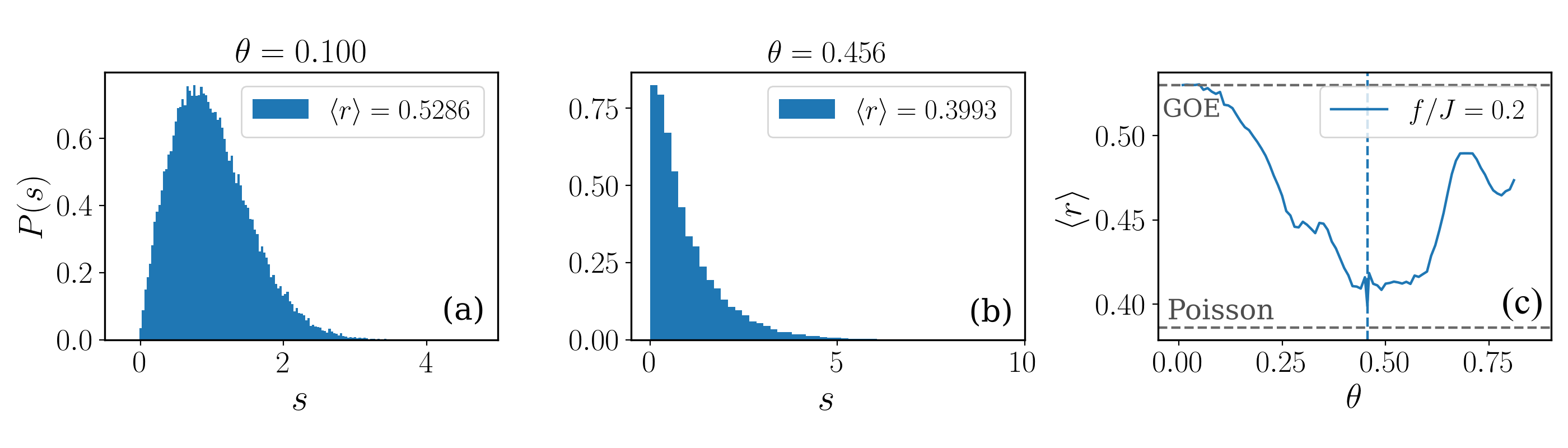}
 	\caption{Level spacing distributions for $\theta=0.1$ and $\theta=0.456$ (close to the ballistic point) are shown in panels (a) and (b) respectively. In (c), variation of mean level spacing ratio $\langle r\rangle$ is plotted  versus $\theta$ showing change in level spacing statistics from GOE to Poissonian at the integrable point $\cos ^{-1}(f/J)/3$.
 	\label{fig:level_spacing}}
 \end{figure} 

\begin{figure}
    \centering
 	\includegraphics[width=0.95\columnwidth]{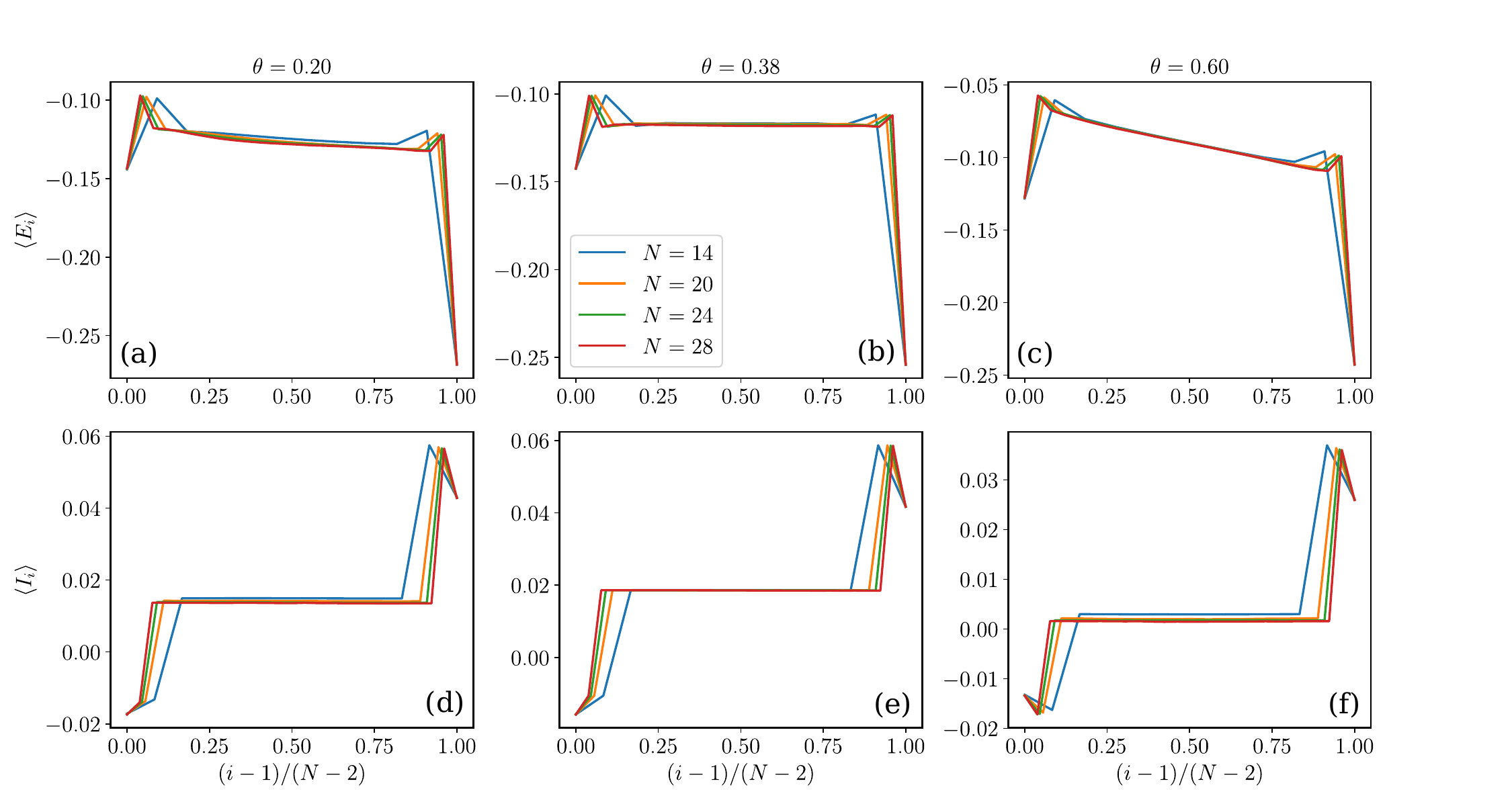}
 	\caption{Spatial profile of energy density and current for system sizes $N$=14, 20, 24, and 28 with position shown on the $x$-axis rescaled by a factor of $1/(N-2)$. Profiles for Hamiltonian parameters $\theta=0.2$ in panel (a) and (d), $\theta=0.38$ in panel (b) and (e), and $\theta=0.2$ in panel (c) and (f). $\theta\sim0.38$ is close to the integrable point. 
 	\label{fig:Profile_ferro}}
 \end{figure} 

We present the estimated $\gamma(\theta)$ as a function of $\theta$ in panels (b) and (c) of Fig. $\ref{fig:Gamma_ferro}$. 
The estimates suggest a clear ballistic energy transport only at the integrable point where $\gamma\approx 0$.
In Panel (b) of Fig. $\ref{fig:Gamma_ferro}$, scaling exponents computed using the two different dissipators show qualitatively the same behavior, and the two estimates quantitatively agree except in a region near small $\theta$. 
We suspect that the difference at the small $\theta$ may be a consequence of different length scales associated with thermalization at the boundary, resulting in different effective lengths for the chain.
In Fig. $\ref{fig:Gamma_ferro}$(c) $\gamma$ is plotted for NESS obtained using the dissipator $\mathcal{D}^0$ for $f/J=0.2$ and 0.4, showing ballistic transport at the expected value of $\theta=\frac{1}{3}\cos^{-1}(f/J)$.

Level spacing statistics (within a symmetry sector of $\mathbb{Z}_3$ parity) computed in a finite system of size $N=11$ (Fig. \ref{fig:level_spacing}) show Poisson statistics at the integrable point and a mixture of GOE and Poisson distributions at other values of $\theta$. The distribution is closer to GOE away from the integrable points. 
Consistent with this, the estimates of $\gamma$ increase away from the integrable points, however it does not indicate fully diffusive behavior in any region of $\theta$.
Studies in disordered spin-1/2 systems have suggested large length scales at weak disorder leading to super-diffusive behavior being observed in finite size calculations \cite{PhysRevB.98.180201,PhysRevLett.117.040601}. We cannot rule out a similar possibility - that a diffusive behavior emerges in larger systems - with the results from the currently accessible system sizes. Spatial profiles of the energy density and current in the NESS for the super-diffusive and ballistic cases are shown in Fig. \ref{fig:Profile_ferro}. As expected the energy density is independent of the position in the bulk in the case of the ballistic system.

\begin{figure}
    \centering
 	\includegraphics[width=0.5\columnwidth]{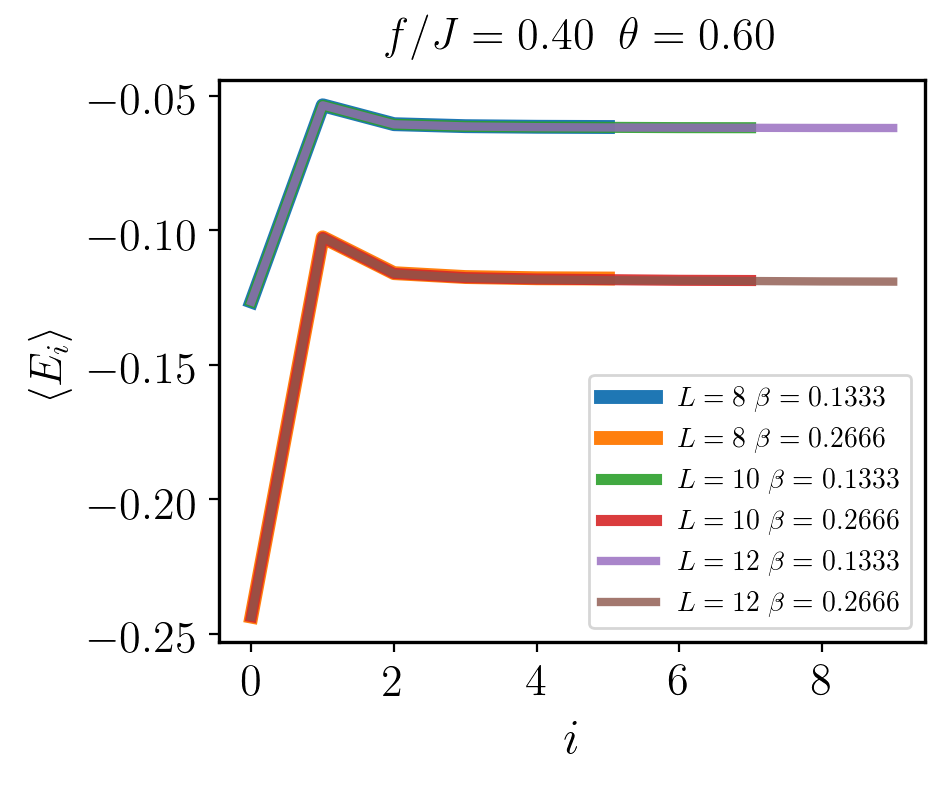}
 	\caption{Energy density as a function of the position in the NESS obtained after attaching only one bath to a chain. The two different lines indicate the energy densities realized upon attaching baths with parameters $\beta_R$ and $\beta_L$. Different overlapping lines of different thicknesses show the data for different system sizes. 	\label{fig:bath_thermalization}}
 \end{figure}

\begin{figure}
    \centering
 	\includegraphics[width=0.8\columnwidth]{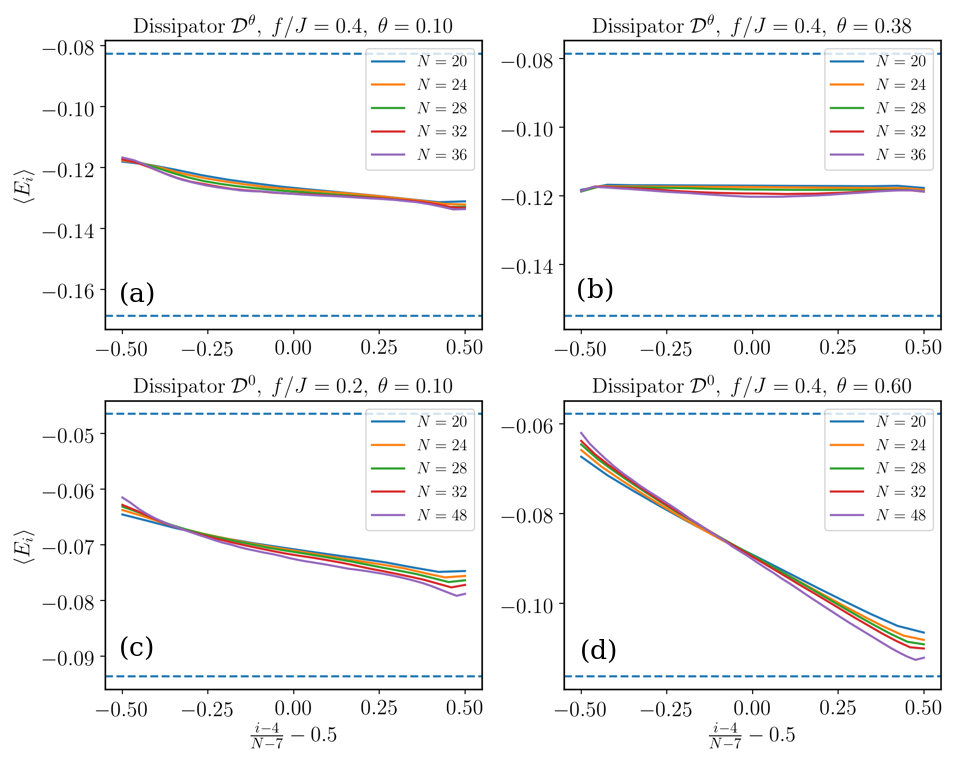}
 	\caption{Each panel shows the energy densities as a function of the position for different system sizes. Position on the x-axis has been rescaled and shifted that center of the chain is at 0 and the $4^{\rm th}$ spin from the ends are at $\pm 0.5$. The two dotted lines show the expected energy densities had the each one of the baths fully equilibrated with the chain (See Fig \ref{fig:bath_thermalization}). The four panels show the data for four different cases. Panels (a) and (c) show results for $\theta$ less than that of the integrable point. Panel (b) shows the data at a $\theta$ very close to the integrable point. Panel (d) shows the same at $\theta$ larger than that of the integrable point.
 	\label{fig:energyprofiles_vs_thermalized_ends}}
 \end{figure}

The analysis in this section relies on the scaling of the current with system size. This yields $\gamma$ provided that the energy densities at the ends of the chain are independent of the system sizes (such that conductance is proportional to the current). In very large systems this can be true, but in small systems the energy densities can be affected by the bath at the other end, resulting in an energy difference that is system size dependent. An estimate of the local energy density that will be realized at the ends if there were local equilibration near the bath can be obtained by attaching only bath to the system. We performed this calculation for each of the two baths. Figure \ref{fig:bath_thermalization} presents the results one of these calculations. 
Figure \ref{fig:energyprofiles_vs_thermalized_ends} shows examples of energy densities as a function of position for different system sizes and parameter regimes (sites very close to the baths have been excluded). The estimates of the expected energy densities if the baths had locally equilibrated with the ends of the chain are shown in dotted lines. 

At the $\theta$ very close to the integrable point (Fig. \ref{fig:energyprofiles_vs_thermalized_ends} panel (b)), the energy densities are midway between the bath energy densities (dotted lines). The energies are approximately independent of the position and system size. In the case of the $\theta$ larger than the integrable value (panel (d) of Fig \ref{fig:energyprofiles_vs_thermalized_ends}), the energy densities realized in the chain are very close to the bath energy densities (indicated by the dotted lines). In the case of $\theta$ smaller than that of the integrable point, the energy densities are position dependent but are far from the estimated bath energy densities. The system size dependence of these energy density difference may then need to be taken into account to make a correct estimate of $\gamma$.

 \begin{figure}
    \centering
 	\includegraphics[width=0.8\columnwidth]{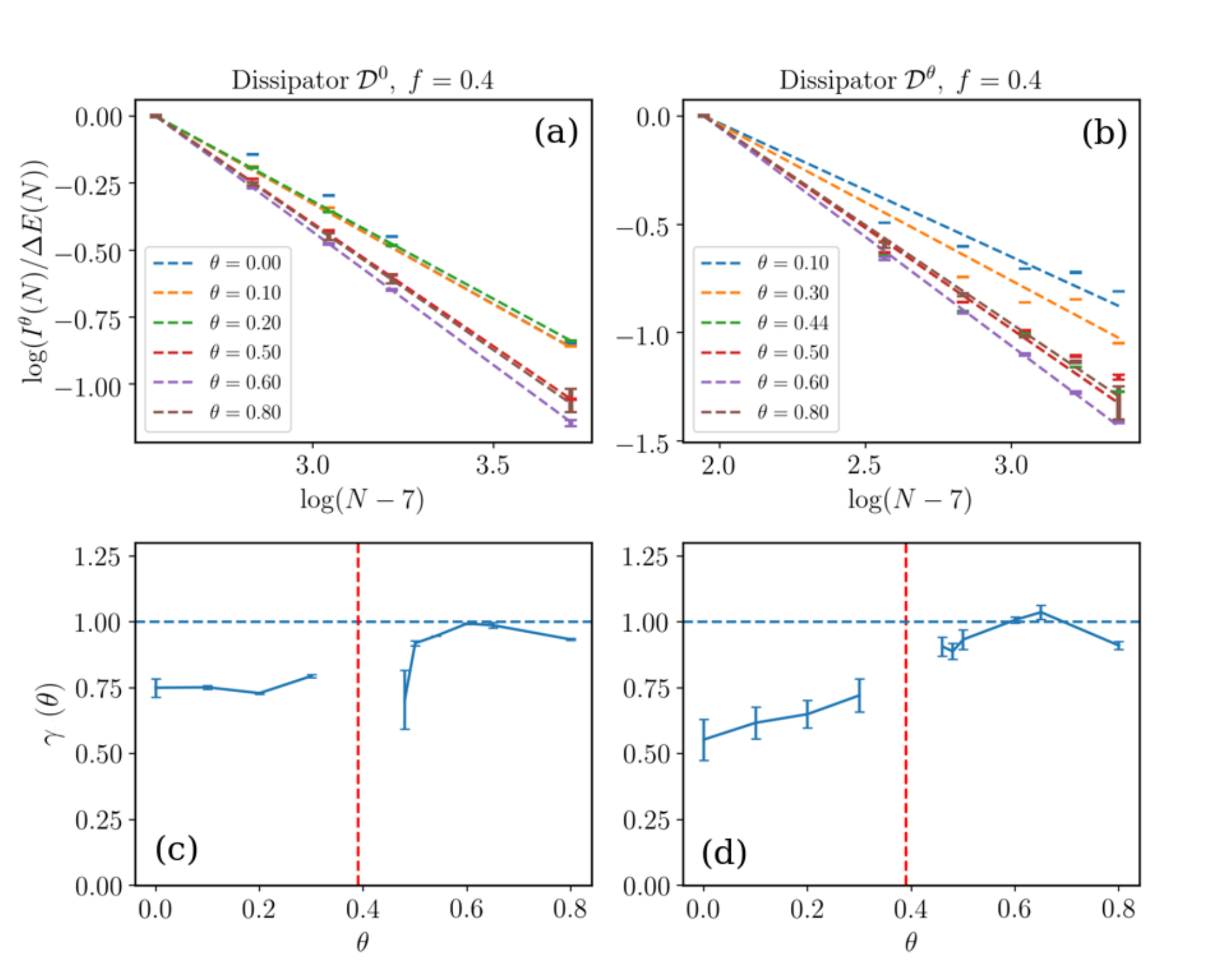}
 	\caption{
In panel (a) and (b), $\log (I^\theta (N)/\Delta E)$ vs $\log (N-7)$ is shown for both dissipators $\mathcal{D}^0$ and $\mathcal{D}^\theta$. Scaling exponent $\gamma$ is obtained by linearly fitting $\log (I^\theta (N)/\Delta E)$ vs $\log (N-7)$ data and is plotted as function of $\theta$ in panels (c) and (d). $\theta$ in the vicinity of integrable points (vertical dashed line) are not shown as the numerically obtained conductance $\kappa$ show wild oscillations due to vanishing energy gradient.  	
 	\label{fig:conductance-scaling}}
 \end{figure}

In Fig. \ref{fig:conductance-scaling} we show the results of the $\gamma$ estimated from the scaling with system size of the conductance. In order to define the conductance, we have assumed that the energy density differences are proportional to temperature differences, taking the ratio of the current to the energy density difference between the $4{\rm th}$ site from either ends of the chain, distance between them being $N-7$. The scaling exponent obtained by fitting the conductance to $N^{-\gamma(\theta)}$ in the panels (c) and (d). The results indicate a larger value of $\gamma$ than what was obtained from scaling of current.

For $\theta$ larger than that of the integrable point, the sites near the ends appear to have nearly equilibrated with the bath (Fig. \ref{fig:energyprofiles_vs_thermalized_ends}(d)). In these cases we find the scaling $\gamma$ to be very close to that of a diffusive system. For smaller $\theta$, where the energy gradients are smaller and much larger system sizes may be needed in order to reliably estimate the true scaling properties. We have not shown the conductance scaling in the vicinity of the integrable points as the energy gradients are nearly zero and numerically estimated conductances show wild variations. 

\begin{figure}
    \centering
 	\includegraphics[width=0.9\columnwidth]{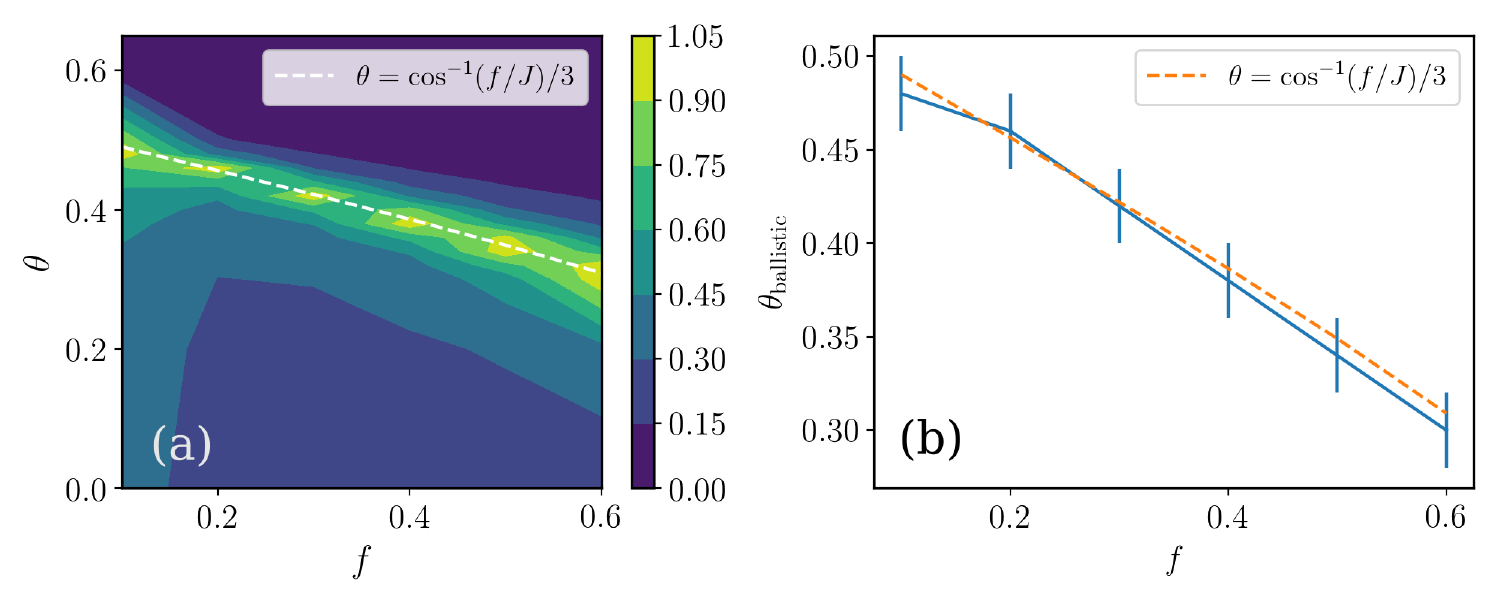}
 	\caption{(a) Rescaled current $I_{\rm{rescaled}}$ of NESS as a function of $\theta$ and $f/J$ is plotted. Comparison of numerically estimated $\theta_{\mathrm{ballistic}}$ and the integrable line $\theta=\cos^{-1}(f/J)/3$ is shown in (b). 
 	\label{fig:conductance_heatmap}}
 \end{figure}

We now discuss a broader range of $f$ values. For not too small system sizes, we expect the peak current and conductance $\kappa$ to occur at the $\theta$ values exhibiting ballistic transport. We may therefore use the peak conductance at each $f$ as a proxy to identify the values of $\theta$ at each $f$ exhibiting ballistic transport.
Figure \ref{fig:conductance_heatmap} shows the estimated current re-scaled and shifted by $f$-dependent constants chosen such that for each $f$, the maximum value of $I_{\rm{rescaled}}$ is 1 and minimum is $0$. Within the numerical uncertainties due to the finite resolution of $\theta$ values, we find that the peak current occurs along the expected line $f/J=\cos(3\theta)$ of integrable points \cite{AUYANG1987219,Fendley2012}.
 
 \subsection{Operator space entanglement}
\begin{figure}
    \centering
 	\includegraphics[width=0.8\columnwidth]{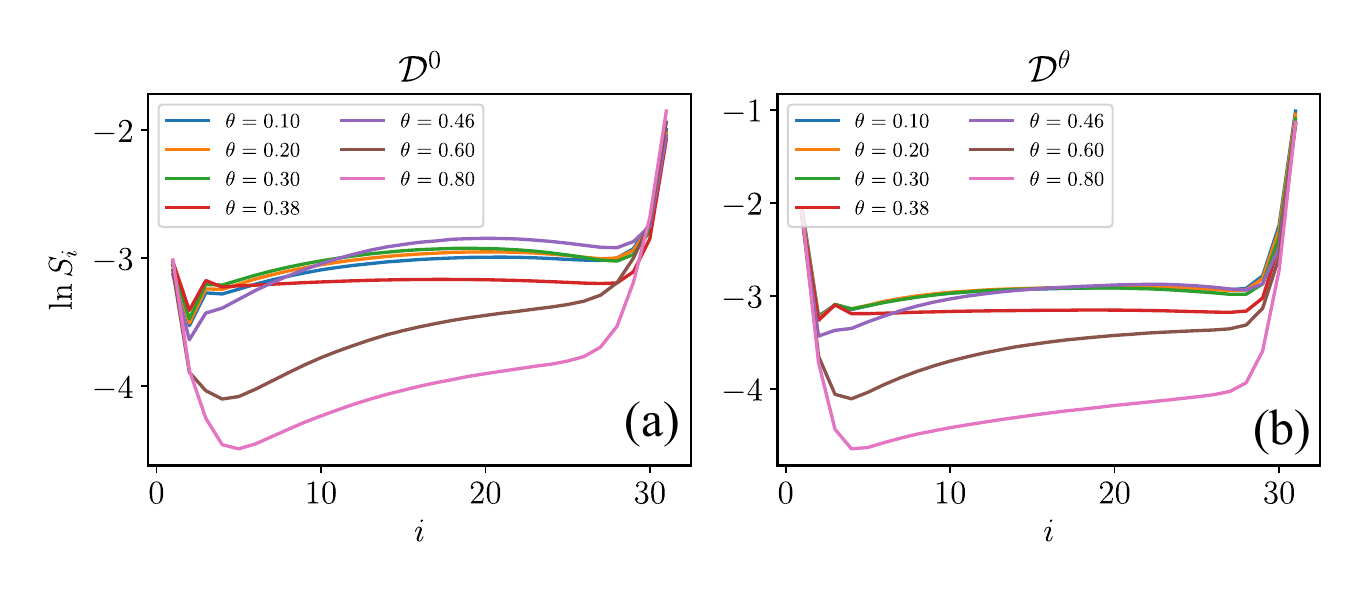}
 	\caption{Operator space entanglement entropy $S_i$ plotted as a function of the bond location $i$. All data are for system size $N=32$ and at $f/J=0.4$. The two panels show the entropy for the NESS obtained under the dissipators $\mathcal{D}^0$ and $\mathcal{D}^\theta$. 
 	\label{fig:OSEE}}
 \end{figure}

Analogous to the notion of entanglement between different bipartitions of many body states, one can define an operator space entanglement entropy (OSEE) \cite{PhysRevA.76.032316,PhysRevB.79.184416} from the MPS representation of the density operator. From the Schmidt decomposition of the state across a partition located at bond $i$, the entropy can be computed as $S_i = -{\rm{Tr}}R_i\log R_i$ where $R_i$ is the reduced density matrix obtained as the partial trace ${\rm Tr}_{j>i}\left| \rho\rangle \langle\rho \right|$.
OSEE of the NESS at different locations of the partition is shown for system size $N=32$ and $f/J=0.4$ is plotted in fig.\ref{fig:OSEE}. 
Empirically we find that at the integrable points, away from the edges, the OSEE is independent of the location of the partition, and for the non-integrable points, $S_i$ shows weak position dependence. 
 \FloatBarrier
The singular values from which the OSEE was constructed also is weakly position dependent in the case of the integrable points. Translation invariance of the entropy as well as of the expectation values of the local operators - energy density and current - at the integrable point suggest the possibility of a translation invariant MPS approximation for the NESS at the integrable points similar to Ref \cite{Znidaric2010}.
 
 \section{Conclusion}\label{Conclusion}
A large body of studies on quantum transport in spin chains performed primarily on spin-half models have indicated that integrable systems show a ballistic energy transport and deviations from integrability generally lead to a diffusive behavior\cite{Saito1996ThermalCI,Saito2,Mej_a_Monasterio_2005} with possible exceptions\cite{PhysRevB.98.235128}. 

In this work we have studied the transport properties of the $\mathbb{Z}_3$ clock model that goes beyond the spin half chains. At the integrable points in the model parameter space, NESS shows a system size independent current, suggesting a ballistic energy transport. At all other values of the parameters the current decreases with the system size. The transport scaling exponent $\gamma$ estimated from scaling of the current alone shows indicates a super-diffusive behavior. Careful analysis of the energy density profiles suggests that this is likely to be a consequence of finite size effects in the system. System size dependence of the energy gradient also needs to be taken into account. The scaling exponent inferred from the conductance instead shows the values closer to diffusive behavior. 
The results demonstrate the connection between integrability and ballistic transport in a larger class of models beyond the well-studied spin half chains.

We have used local Lindblad coupling to the edges of a finite chain of chiral $\mathbb{Z}_3$ clock to approximately model the coupling of the system to the bath. Within this approach, we obtained similar results when different dissipator models were used at the edge, suggesting a robustness of the results to the precise nature of the coupling of the system to the bath. Direct computation of the Drude weights can be an independent approach to verify the characterization of transport properties in the model \cite{Kubo1,Kubo2,Subroto1,MooreDrudeWt,KarraschDrudeWt}.

\section{Acknowledgement}
We thank F Alet, S L Srivastava, A Trivedi for the useful discussions and A Thalapillil for sharing computational resources.
NN would like to thank the organizers of the ``Bangalore School on Statistical Physics XII'' (code: ICTS/bssp2021/6) for giving the opportunity to be a part of their workshop and learn about open quantum system. 
Calculations were performed on codes built using the ITensor Library\cite{itensor}.
SGJ acknowledges DST/SERB Grant No. ECR/2018/001781 and CNRS-IISER Pune joint grant for providing funding. We also thank National Supercomputing Mission (NSM) for providing computing resources of ‘PARAM Brahma’ at IISER Pune, which is implemented by C-DAC and supported by the Ministry of Electronics and Information Technology (MeitY) and Department of Science and Technology (DST), Government of India.

 \FloatBarrier

\section*{References}
\bibliography{Z3energy}

\end{document}